\documentclass[a4paper,11pt]{article}
\usepackage{pos,slashed}

\title{The detection of cosmic neutrino background with helicity-changing decays}
\ShortTitle{The detection of C$\nu$B with helicity-changing decays}

\author*[a,b]{Jihong Huang}
\author[a,b]{Shun Zhou}

\affiliation[a]{Institute of High Energy Physics, Chinese Academy of Sciences, Beijing 100049, China}

\affiliation[b]{School of Physical Sciences, University of Chinese Academy of Sciences, Beijing 100049, China}

\emailAdd{huangjh@ihep.ac.cn}
\emailAdd{zhoush@ihep.ac.cn}

\abstract{In this talk, we present the investigation of the invisible decays of a heavy massive neutrino into a lighter neutrino and a massless Nambu-Goldstone boson, i.e., $\nu_i^{} \to \nu_j^{} + \phi$. The total decay rates are calculated in the most general case, where the individual helicities of both parent and daughter neutrinos are specified. We then examine the evolution of the number densities of cosmological relic neutrinos throughout the expansion of the Universe, and explore the consequent impacts on the capture rates in PTOLEMY-like experiments. The total event rates can be significantly modified compared to those in the scenario of stable neutrinos, with helicity-changing decays playing an especially important role in the Dirac neutrino case.}

\FullConference{19th International Conference on Topics in Astroparticle and Underground Physics (TAUP2025)\\
24–30 Aug 2025\\
Xichang, China\\}

\begin{document}
\maketitle

\section{Introduction}

Neutrino oscillation experiments have revealed that at least two species of neutrinos are massive, and it will be interesting to investigate whether a heavier neutrino can decay into a lighter one and other elementary particles within or beyond the Standard Model (SM). In the Majoron model for explaining the neutrino mass origin, massive neutrinos can interact with the Nambu-Goldstone boson $\phi$, i.e., the Majoron, after the spontaneous breaking of the global lepton number symmetry, which is described by
\begin{eqnarray}
	\label{eq:L_D}
	{\cal L}^{}_{\rm D} = \sum_i \left(\overline{\nu^{}_i} {\rm i}\slashed{\partial} \nu^{}_i - m^{}_i \overline{\nu^{}_i} \nu^{}_i\right) + \frac{1}{2} \partial^{}_\mu \phi \partial^\mu \phi - \left[ {\rm i}\phi \sum_{i, j} g^{}_{ij} \overline{\nu^{}_i} \gamma^5 \nu^{}_j + {\rm h.c.} \right] 
\end{eqnarray}
for Dirac neutrinos. The interaction term and its Hermitian conjugate in the square brackets induce the direct decays of antineutrinos $\overline{\nu_i^{}} \to \overline{\nu_j^{}} +\phi$ and neutrinos $\nu_i^{} \to \nu_j^{} +\phi$, respectively. For Majorana neutrinos, we similarly have 
\begin{eqnarray} 
	\label{eq:L_M}
	{\cal L}^{}_{\rm M} = \frac{1}{2} \sum_i \left(\overline{\nu^{}_i} {\rm i}\slashed{\partial} \nu^{}_i - m^{}_i \overline{\nu^{}_i} \nu^{}_i \right) + \frac{1}{2} \partial^{}_\mu \phi \partial^\mu \phi - \left[ {\rm i}\phi \sum_{i, j} g^{}_{ij} \overline{\nu^{}_i} \gamma^5 \nu^{}_j + {\rm h.c.} \right] \;, 
\end{eqnarray}
in which two identical interaction terms suggest the decay amplitudes being twice that for the Dirac one. Such invisible decays would have significant impacts on the evolution of the Universe and can be tested in PTOLEMY-like experiments~\cite{Betts:2013uya}, which aim to detect the cosmic neutrino background (C$\nu$B) by capturing relic neutrinos on the beta-decaying nuclei, such as $\nu_e^{} + {}^3{\rm H} \to {}^3{\rm He} + e^- $~\cite{Weinberg:1962zza}. With $100~{\rm g}$ tritium as the target, the capture rate for Majorana neutrinos is $\Gamma_{\rm C\nu B}^{\rm M} \approx 8~{\rm yr}^{-1}$, while that for Dirac neutrinos is $\Gamma_{\rm C\nu B}^{\rm D} \approx 4~{\rm yr}^{-1}$ and can increase to $7~{\rm yr}^{-1}$ in the NO case if the neutrino momentum distribution is taken into account~\cite{capture_rates}. As the capture rates of relic neutrinos involve their specific helicities, neutrino decays with different helicities for parent and daughter neutrinos should be carefully examined, and their impacts on the total capture rates cannot be neglected.

\section{Neutrino Invisible Decays}

From the Lagrangian in Eq.~(\ref{eq:L_D}) or Eq.~(\ref{eq:L_M}), one can directly write down the decay amplitudes for $\nu_i^{}(p_i^{},h_i^{}) \to \nu_j^{}(p_j^{},h_j^{}) + \phi(k)$, with $p_i^{} = (E_i^{},{\bf p}_i^{})$, $p_j^{} = (E_j^{},{\bf p}_j^{})$ and $k=(\omega,{\bf k})$ being the four-momenta of parent and daughter particles. To compute the total helicity-preserving ($h^{}_i = h^{}_j = \pm 1$) and helicity-changing ($h^{}_i = \pm 1$, $h^{}_j = \mp 1$) decay rates, we need the amplitude squared with specific helicities for parent and daughter neutrinos and integrate over the final-state phase space. For Majorana neutrinos, with the velocity of the parent neutrino $\beta^{}_i \equiv \left|{\bf p}_i^{}\right|/E_i^{}$, we arrive at
\begin{eqnarray}
    \Gamma^{\rm M}_{\pm\pm,ij} &=& \frac{g^2_{ij}m^{}_i}{4\pi} \sqrt{1 - \beta^2_i} \left\{ \frac{1}{2} \left(1 - r^2_{ji}\right) (1 - r^{}_{ji})^2 - r^{}_{ji} (1 - r^2_{ji}) - 2r^2_{ji} \ln r^{}_{ji} \right. \nonumber \\
	&& \left. - (\beta^{-2}_i - 1) \left[ (1+r^{}_{ji}+r^2_{ji}) r^{}_{ji} \ln r^{}_{ji} + \frac{1}{4} (1 - r^2_{ji}) (1 + 4r^{}_{ji} + r^2_{ji})\right]\right\} \; , \label{eq:GammaMajpresG} \\
	\Gamma^{\rm M}_{\pm\mp,ij} &=& \frac{g^2_{ij}m^{}_i}{4\pi} \sqrt{1 - \beta^2_i} \left\{ \frac{1}{2} \left(1 - r^2_{ji}\right) (1 - r^{}_{ji})^2 + r^{}_{ji} (1 - r^2_{ji}) + 2r^2_{ji} \ln r^{}_{ji} \right. \nonumber \\
	&& \left. + (\beta^{-2}_i - 1) \left[ (1+r^{}_{ji}+r^2_{ji}) r^{}_{ji} \ln r^{}_{ji} + \frac{1}{4} (1 - r^2_{ji}) (1 + 4r^{}_{ji} + r^2_{ji})\right] \right\} \; , \label{eq:GammaMajflipG}
	\end{eqnarray}
for $\beta^{}_i > \beta^*_{ji} \equiv (1 - r^2_{ji})/(1 + r^2_{ji})$; and 
\begin{eqnarray}
	\Gamma^{\rm M}_{\pm\pm,ij} &=& \frac{g^2_{ij}m^{}_i}{4\pi} \sqrt{1 - \beta^2_i} \left\{ \frac{1}{2} \left(1 - r^2_{ji}\right) (1 - r^{}_{ji})^2 - (1+r^{}_{ji}+r^2_{ji}) r^{}_{ji} \beta^{-1}_i \right. \nonumber \\
	&& \left. + \left[ r^{}_{ji} + \frac{1}{2} (\beta^{-2}_i - 1) (1+r^{}_{ji}+r^2_{ji})\right]  r^{}_{ji} \ln \left(\frac{1+\beta^{}_i}{1-\beta^{}_i}\right) \right\} \; , \label{eq:GammaMajpresL} \\
	\Gamma^{\rm M}_{\pm\mp,ij} &=& \frac{g^2_{ij}m^{}_i}{4\pi} \sqrt{1 - \beta^2_i} \left\{ \frac{1}{2} \left(1 - r^2_{ji}\right) (1 - r^{}_{ji})^2 + (1+r^{}_{ji}+r^2_{ji}) r^{}_{ji} \beta^{-1}_i \right. \nonumber \\
	&& \left. - \left[ r^{}_{ji} + \frac{1}{2} (\beta^{-2}_i - 1) (1+r^{}_{ji}+r^2_{ji})\right]  r^{}_{ji} \ln \left(\frac{1+\beta^{}_i}{1-\beta^{}_i}\right) \right\} \; , \label{eq:GammaMajflipL}
	\end{eqnarray}
for $\beta^{}_i \leqslant \beta^*_{ji} \equiv (1 - r^2_{ji})/(1 + r^2_{ji})$, with the neutrino mass ratio $r^{}_{ji} \equiv m^{}_j /m^{}_i$. The total decay rate for $\nu_i^{}(p_i^{},h_i^{})$ can be defined as $\Gamma_{\pm,i}^{\rm M} \equiv \sum_j \Gamma^{\rm M}_{\pm\pm,ij} + \Gamma^{\rm M}_{\pm \mp,ij}$ by summing over all the possible decay channels with daughter neutrinos $\nu_j^{}(p_j^{},h_j^{})$. For Dirac neutrinos, the decay rates can be similarly obtained and the relations $\Gamma^{\rm D}_{h_i^{}h_j^{},ij} = \Gamma^{\rm M}_{h_i^{}h_j^{},ij} /4$ hold. The critical velocity for parent neutrinos $\beta^{*}_{ji}$ is determined by whether the direction of the daughter neutrino momentum is perpendicular to that of the parent neutrino. These two configurations of angular distributions make the decay rates in two cases quite different. By adopting results from the latest global analysis of neutrino oscillation data~\cite{Esteban:2024eli} and assuming the lightest neutrino mass of $0.1~{\rm meV}$, the asymmetry between the helicity-preserving and -changing decay rates can reach up to $20\%$ in the case of normal mass ordering (NO) and as large as $40\%$ in the case of inverted mass ordering (IO). This further highlights the necessity of studying neutrino decays in which their helicities are changed.

\section{Cosmic Neutrino Background}

Neutrinos decoupled from the thermal bath at the redshift $z_{\rm d}^{} \approx 10^{10}$ in the standard cosmology, corresponding to the temperature $T \approx 1~{\rm MeV}$. Afterwards, the three-momentum ${\bf p}(z)$ and the temperature $T_\nu^{}(z)$ evolve with the redshift according to $|{\bf p}(z)| = |{\bf p}(z_{\rm d}^{})|(1+z)/(1+z_{\rm d}^{})$ and $T_\nu^{}(z) = T_\nu^{}(z_{\rm d}^{})(1+z)/(1+z_{\rm d}^{})$. The corresponding number density is given by $\overline{n} (z) = 3\zeta(3)T^3_\nu(z)/(4\pi^2)$, and the spectrum follows the modified Fermi-Dirac (FD) distribution
\begin{eqnarray}\label{eq:FD}
	f_{\rm FD}^{} \left[|{\bf p}(z)|, T_\nu^{}(z) \right] = \frac{1}{\exp \left[|{\bf p}(z)|/T_\nu^{}(z)\right] + 1} \;. 
\end{eqnarray} 
We further define the modified number density as $n_i^{}(z) \equiv \overline{n}_i^{}(z) a(z)^3 $, where the effect from the expansion of the Universe is absorbed into the scale factor $a(z) \equiv 1/(1+z)$. When taking the neutrino decays into account, the number densities of parent and daughter neutrinos with the same helicity at the redshift $z$ should be calculated as
\begin{eqnarray} 
	\label{eq:daughter_number_modified}
	n_i^{\pm} (z) &=& \frac{\displaystyle \int_{0}^{\infty} n_0^{} {\rm e}^{-\lambda_i^{} (z)}_{} \; f_{\rm FD}^{} [|{\bf p}^{}_i|,T_\nu^{}(z)] |{\bf p}^{}_i|^2\ {\rm d}|{\bf p}^{}_i|}{\displaystyle \int_{0}^{\infty} f_{\rm FD}^{} [|{\bf p}^{}_i|,T_\nu^{}(z)] |{\bf p}^{}_i|^2\ {\rm d}|{\bf p}^{}_i|} \;, \nonumber \\
	n_j^{\pm}(z) &=& \frac{\displaystyle \int_{0}^{\infty} n_0^{} \left[1 - {\rm e}^{-\lambda_i^{} (z)}_{} \right] {\cal B}^{\rm M+}_{ij} \; f_{\rm FD}^{} [|{\bf p}^{}_j|,T_\nu^{}(z)] |{\bf p}^{}_j|^2\ {\rm d}|{\bf p}^{}_j|}{\displaystyle \int_{0}^{\infty} f_{\rm FD}^{} [|{\bf p}^{}_j|,T_\nu^{}(z)] |{\bf p}^{}_j|^2\ {\rm d}|{\bf p}^{}_j|} \;.
\end{eqnarray}
Here the FD spectra of $\nu_{i,j}^{}$ have been included, ${\cal B}^{\rm M+}_{ij}\equiv \Gamma^{\rm M}_{\pm\pm,ij}/\Gamma^{\rm M}_{\pm,i}$ is the branching ratio of the helicity-preserving decay channel, while that for the helicity-changing decay can be similarly defined as ${\cal B}^{\rm M-}_{ij}\equiv \Gamma^{\rm M}_{\pm\mp,ij}/\Gamma^{\rm M}_{\pm,i}$. The suppression factor $\lambda_i^{} (z) $ is calculated by
\begin{eqnarray} 
	\label{eq:lambda}
	\lambda_i^{} (z) = \int_z^{z_i^{\rm decay}} \frac{{\rm d} z^\prime \ \Gamma_{\pm,i}^{{\rm M} *}}{(1+z^\prime) H(z^\prime) \gamma(z^\prime)} \;,
\end{eqnarray}
with the Lorentz factor $\gamma^{}_i(z) \equiv 1/\sqrt{1-\beta_i^{2}(z)}$, the Hubble parameter $H(z)$ and the total decay rate for $\nu_i^{}$ in the rest frame $\Gamma_{\pm,i}^{{\rm M} *}$. The redshift $z_i^{\rm decay}$ describes the moment when a substantial fraction of $\nu^{}_i$ starts to decay for a given coupling constant. For the second-heaviest neutrinos, we should also take their secondary decays into account, where the momentum distribution functions will also be modified. See Sec.~3.1 of Ref.~\cite{Huang:2024tbo} for further details.

\begin{figure}[t]
	\centering
	\includegraphics[scale=0.45]{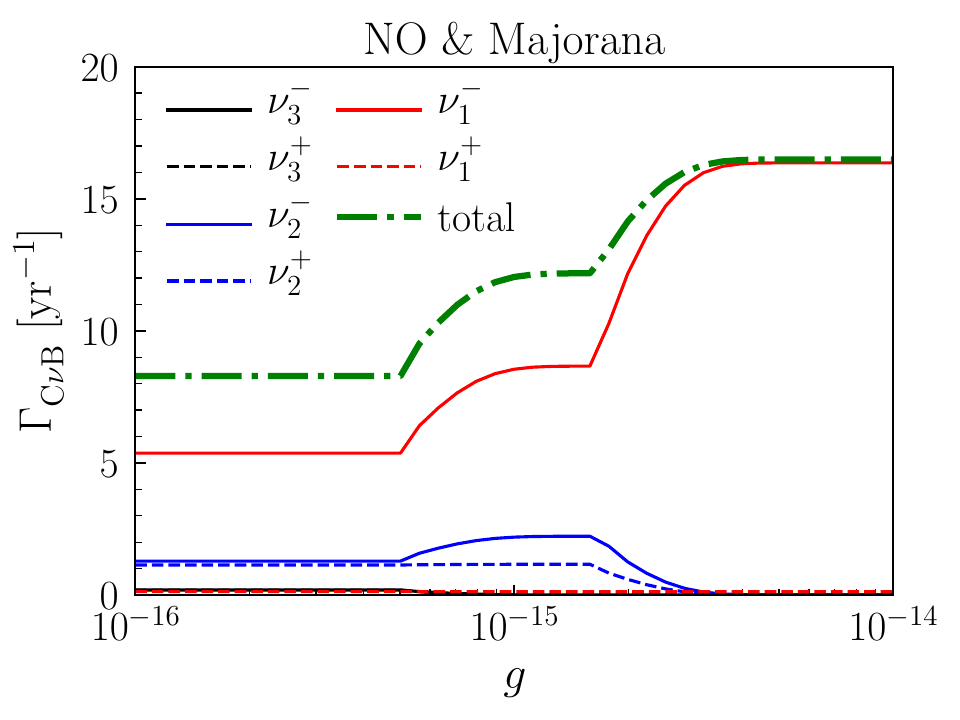}
	\hspace{-0.5cm}
	\includegraphics[scale=0.45]{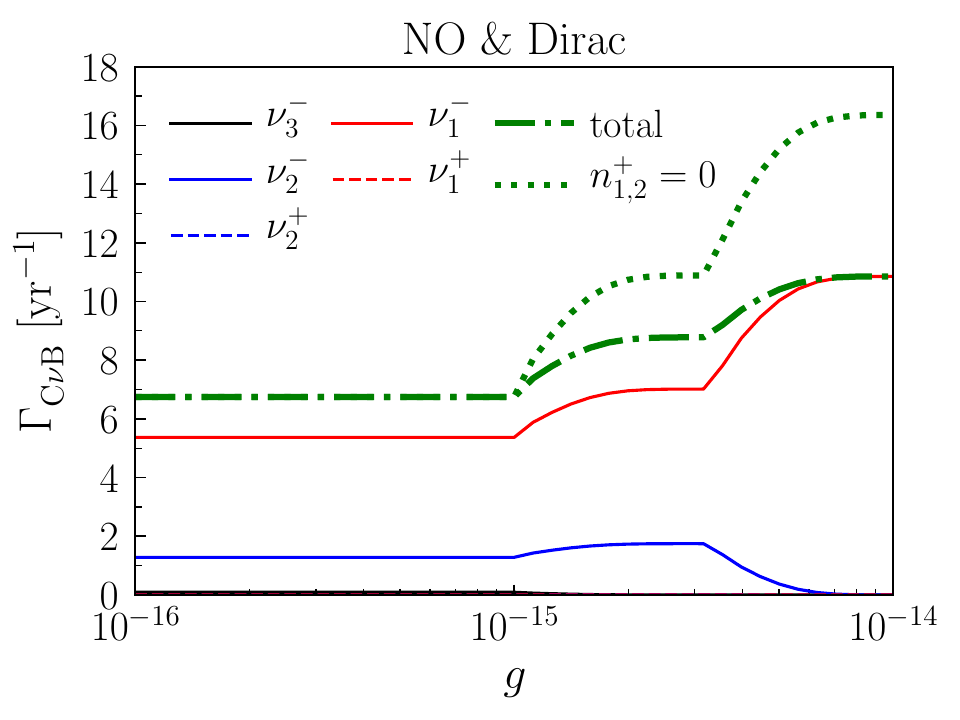} 
	\vspace{-0.3cm}
	\caption{The capture rates for Majorana (left) and Dirac (right) neutrinos in the case of NO~\cite{Huang:2024tbo}. The green dot-dashed curves represent the total capture rates, while the contributions from each mass eigenstate with a specific helicity are shown separately. For Dirac neutrinos, the capture rates with all decays being helicity-preserving are plotted in the green dotted curve.}
	\label{fig:cap_NO}
\end{figure}

After determining the evolution of neutrino number densities for different mass eigenstates with specific helicities, we arrive at their present values by setting the redshift $z=0$. They are used to calculate the total capture rates in PTOLEMY-like experiments~\cite{capture_rates}
\begin{eqnarray} \label{eq:capture_rate}
	\Gamma_{\rm C\nu B}^{} = N_{\rm T}^{} \overline{\sigma} \sum_{i=1}^3 \sum_{h_i^{}=\pm 1} \left|U_{ei}^{}\right|^2 n_i^{}(h_i^{}) {\cal A}(\beta_i^{},h_i^{}) \;,
\end{eqnarray}
with the number of tritium nuclei $N_{\rm T}^{} \approx 2\times 10^{25}$, and the cross section $\left|U_{ei}^{}\right|^2 \overline{\sigma}$ where $\overline{\sigma} \approx 3.834 \times 10^{-45}~{\rm cm}^2$ and $U$ is the leptonic flavor mixing matrix. The function ${\cal A}(\beta_i^{},h_i^{}) \equiv 1-h_i^{} \left<\beta_i^{}\right> $ depends on the helicity of captured neutrinos with the number density $n_i^{}(h_i^{})$ and their average velocity $\left<\beta_i^{}\right>$. The numerical results in the NO case for various coupling constants are presented in Fig.~\ref{fig:cap_NO} as green dot-dashed curves, which are obtained with the lightest neutrino mass being $0.1~{\rm meV}$ and other parameters adopted from Ref.~\cite{Esteban:2024eli}. All the coupling constants are set to be equal, i.e., $g_{ij}^{} \equiv g$, taken in the range $g\in [10^{-16}, 10^{-10}]$ to satisfy the most restrictive constraint from cosmological observations~\cite{Wong}. Capture rates in the standard case are reproduced for small coupling constants, while for larger ones they could reach $\Gamma_{\rm C\nu B}^{\rm M} \approx 16~{\rm yr}^{-1}$ for Majorana neutrinos. Two step-like increases come from the contributions of $\nu_3^{}$ and $\nu_2^{}$ decays, respectively. For Dirac neutrinos, the total rate is $\Gamma_{\rm C\nu B}^{\rm D} \approx 11~{\rm yr}^{-1}$ since only neutrinos can be captured in this case. In order to emphasize the effects from the helicity-changing decays, we also plot the capture rates by assuming all decays are helicity-preserving as the green dotted curve. There is no change for Majorana neutrinos, while the total rates for Dirac neutrinos increase to $\Gamma_{\rm C\nu B}^{\rm D} \approx 16~{\rm yr}^{-1}$ since the number densities of left-helical neutrinos will be much larger now. In the IO case, the decay rates of $\nu_2^{} \to \nu_1^{} + \phi$ are highly suppressed, and the small matrix element $\left|U_{e 3}^{}\right|^2 \approx 0.023$ results in the capture rate as low as $1~{\rm yr}^{-1}$ for both Majorana and Dirac cases. Meanwhile, there is also no significant difference when only considering the helicity-preserving decays.

\section{Summary}

We study the invisible decays of massive neutrinos and their implications for the ${\rm C\nu B}$ detection. The total decay rates in both the helicity-preserving and helicity-changing decay channels are calculated, and significant asymmetries between them have also been noticed. The number densities of cosmic relic neutrinos are derived during the expansion of the Universe. By varying the coupling constant between massive neutrinos and the Nambu-Goldstone boson, we evaluate the capture rates of ${\rm C\nu B}$ in PTOLEMY-like experiments, which can be as large as $16~{\rm yr}^{-1}$ for Majorana neutrinos and $11~{\rm yr}^{-1}$ for Dirac neutrinos in the case of NO, while less than $1~{\rm yr}^{-1}$ in the case of IO. In the future, PTOLEMY-like experiments will be capable of measuring the absolute neutrino mass scale and detecting cosmic relic neutrinos, thereby establishing themselves as a unique platform to explore the intrinsic properties of massive neutrinos and test theories beyond the SM.\\


{\it This work was supported by the National Natural Science Foundation of China under grant No.~11835013 and No.~12475113, by the CAS Project for Young Scientists in Basic Research (YSBR-099), and by the Scientific and Technological Innovation Program of IHEP under grant No.~E55457U2.}

\end{document}